# Unpacking Musical Symbolism in Online Communities: Content-Based and Network-Centric Approaches

Kajwan Ziaoddini[†]
Ethnomusicology
University of Maryland
College Park, USA
kajwan@umd.edu

## ABSTRACT

This paper examines how musical symbolism is produced and circulated in online communities by combining content-based music analysis with a lightweight network perspective on lyrics. Using a curated corpus of 275 chart-topping songs enriched with audio descriptors (energy, danceability, loudness, liveness, valence, acousticness, speechiness, popularity) and full lyric transcripts, we build a reproducible pipeline that (i) quantifies temporal trends in sonic attributes, (ii) model's lexical salience and co-occurrence, and (iii) profiles mood by genre. We find a decade-long decline in energy (79 → 58) alongside a rise in danceability (59 → 73); valence peaks in 2013 (63) and dips in 2014–2016 (42) before partially recovering. Correlation analysis shows strong coupling of energy with loudness ($r = 0.74$) and negative associations for acousticness with both energy ($r = −0.54$) and loudness ($r = −0.51$); danceability is largely orthogonal to other features ($|r| < 0.20$). Lyric tokenization (>114k tokens) reveals a pronoun-centric lexicon "I/you/me/my" and a dense co-occurrence structure in which interpersonal address anchors mainstream narratives. Mood differs systematically by style: R&B exhibits the highest mean valence (96), followed by K-Pop/Pop (77) and Indie/Pop (70), whereas Latin/Reggaeton is lower (37) despite high danceability. Read through a subcultural identity lens, these patterns suggest the mainstreaming of previously peripheral codes and a commercial preference for relaxed yet rhythmically engaging productions that sustain collective participation without maximal intensity. Methodologically, we contribute an integrated MIR-plus-network workflow spanning summary statistics, correlation structure, lexical co-occurrence matrices, and genre-wise mood profiling that is robust to modality sparsity and suitable for socially aware recommendation or community-level diffusion studies.

## 1. INTRODUCTION

Music is not merely an aesthetic experience but a social code that writes identity, values, and belonging. Music listening, sharing, and talking in virtual communities move beyond entertainment to function as methods of cultural signaling and symbolic communication. Genres, lyrics, and even sound characteristics can be utilized as markers of subcultural membership, political allegiance, or emotional state [1]. These symbolic musical dimensions are increasingly apparent in digital environments like streaming services, fan sites, and social networks, where consumers all together build narratives about artists, styles, and songs [2].

Prior research has come at music analysis from two prevalent ways of thinking. Content-based strategies emphasize the inherent natures of music itself, such as audio characteristics and words. These approaches have also been supplemented with computational linguistics and music information retrieval advances, which are now capable of supporting large-scale song structure, emotional tone, and thematic pattern analysis [3,4]. Network-based approaches, however, study the relational patterns engendered by music-sharing activities, playlist co-occurrence, and artist-fan networks. Such analyses reveal how musical preferences come to be embodied in rich social networks, reveals clustering patterns, influence, and identity formation [5].

While both views have been valuable, they have sometimes been applied apart. Apart from each other, they restrict the potential to grasp how the symbolic meaning of music arises collectively from its formal properties and from its formal spread throughout communities [6]. For instance, a song's lyrical sentiment can touch the listener in another manner based on the position the listener occupies within a web of communities, and genre-based identities can be strengthened through formal repetition among sets of spectators of specific audio properties [6,7].

This paper fills this gap through both content-based and network-based methods of analyzing musical symbolism in social media. With a diverse corpus of popular songs enriched by audio descriptions and transcribed lyrics, we explore (1) semantic and emotional patterns in lyrics, (2) statistical dependencies between audio features, and (3) structural features of user-tagged genre and playlist networks. By the crossing of these dimensions, we seek to reveal how particular genres inscribe subcultural identities, how lyrical messages and sonority signifiers recreate symbolic meanings and repeat them, and how they are transmitted by digital interaction networks. The contributions of this paper are threefold:
1. We introduce a methodology combining MIR-based content analysis and network science measures for studying symbolic music communication.
2. We provide empirical evidence substantiating musical qualities to subcultural identity expression patterns online.
3. We provide theoretical explanations bridging computational research and cultural analyses of music and identity.

By placing computational research in a cultural frame, this work not only contributes to interdisciplinary music studies but also fully explains how digital environments mediate and reconfigure music's social life.

## 2. LITTERATURE REVIEW

Williams et al. [7] explored how niche TikTok groups construct identities, revealing themes like objectification, romanticisation, and trivialisation. They showed that online narratives can actively shape symbolic meanings and group affiliations, a process comparable to our analysis of musical symbolism through content attributes and network dynamics.

Ruebottom et al. [8] investigated how organizations engage heterogeneous communities through open and protective participation work. They demonstrated that these approaches produce different social-symbolic effects, influencing both organizational flexibility and community inequalities. This insight aligns with our study's focus on how symbolic practices shape participation and identity within online musical communities.

Wang et al. [9] conducted a qualitative content analysis of China's Qinglang Action, a regulatory move targeting celebrity reality television. Their study revealed the policy's socio-political and economic implications, highlighting tensions between fandom culture, market dynamics, and state governance. These insights contribute to understanding how state interventions shape digital cultural expressions and community behaviors.

Oliva-Bulpitt et al. [10] presents a neural network approach for Optical Music Recognition that integrates Monte Carlo Dropout to enhance staff-region detection accuracy. By employing multiple non-deterministic predictions and novel combination strategies, the method reduced relative error by 63.6% and improved accuracy by 32.1% over existing techniques. This demonstrates the potential of probabilistic inference in improving robustness for layout analysis tasks. Kozinets [11] introduces "netnography," an adaptation of ethnography for studying online communities to gain consumer insights. The method is faster, simpler, and more cost-effective than traditional ethnography, while remaining naturalistic and unobtrusive. It offers guidelines for rigorous and ethical research, exemplified through a netnography of an online coffee newsgroup and its marketing implications.

Brevik [12] examines extreme outliers in second language (L2) acquisition among Norwegian adolescents, focusing on their English proficiency despite weaker first language (L1) reading skills. Through mixed methods, the study identifies three profiles—Gamers, Surfers, and Social Media Users—whose high engagement with English technology, media, and online activities outside school significantly contributed to their L2 competence.

Casey et al. [13] discusses the rapid shift from physical to digital music formats, driven by the growth of online streaming and downloading services. With music collections reaching tens of millions of tracks, content-based music information retrieval (CBMIR) has become crucial for efficient searching, retrieval, and organization. The paper reviews state-of-the-art CBMIR techniques, including audio cue analysis and symbolic representation, while outlining key challenges such as scalability and interdisciplinary integration for future development.

## 3. MODEL FRAMEWORK
### 3.1. DATASET SELECTION AND ETHICS

We worked with the Spotify 2010–2020 Top 25 Songs and Lyrics database, an open database of 275 songs that hold the top 25 hit songs of every year from 2010–2020 [15]. There is one record per song with song title, artist, genre, release date, audio descriptors (energy, danceability, loudness, liveness, valence, acousticness, speechiness and popularity), and complete lyric transcript. This one building can house multimodal analysis: audio features can be paired up directly with text content without clunky joints. The dataset is leveraged heavily in music analytics research [16], and the presence of valence and danceability features aligns with larger datasets like the Mendeley Lyrics Dataset [17]. Due to copyright issues, only derived features (statistics and embeddings) were saved; raw audio and lyrics files were not released. User identifiers were not collected, thus no direct privacy concerns.

Although representative of mass pop culture, the dataset has its limits: it is grounded in chart hits, mainly in English, and thus partly under-represents alternative scenes and languages other than Western nations. To explore subcultural dynamics, results are interpreted in a guarded manner and informed by literature linking musical styles to subcultural identity.

### 3.2. TEXT PROCESSING

Lyrics were lowercased and tokenised by a regular expression that took out alphabetic sequences which were contiguous.

Technically, the token collection of document d is "tokens" $(d) = \{w \in \Sigma^+ : w$ is a sequence of letters in " d$\}$, where $\Sigma$ consists of the English alphabet. Tokens were counted throughout the corpus: the occurrences of word w are $C(w) = \sum_{d \in D} f_{w,d}$, where $f_{w,d}$ denotes the occurrences of w in document d. Even though regular stopwords were not removed, how frequently pronouns are used in abundance is the concern under investigation (narratives in themselves carry significance). The size of vocabulary, frequencies and relative ratios of top-words were calculated. While more sophisticated methods like TF–IDF or metaphor identification can unearth deeper semantics[5], current analysis centers on frequency and co-occurrence patterns to investigate repeated words as symbols.

### 3.3 AUDIO FEATURE ANALYSIS

For each continuous audio variable $x$ (energy, danceability, loudness etc.), summary statistics were computed: mean $\bar{x} = \frac{1}{n}\sum_i x_i$, standard deviation $\sigma_x = \sqrt{\frac{1}{n-1}\sum_i (x_i - \bar{x})^2}$, and extrema. Temporal trends were assessed by grouping songs by release year and calculating annual means. Pearson's correlation coefficients

between pairs of variables were computed to explore relationships (e.g. between energy and danceability):

$$r_{xy} = \frac{\sum_i (x_i - \bar{x})(y_i - \bar{y})}{(n-1)\sigma_x \sigma_y}. \quad (1)$$

## 3.4 CO-OCCURRENCE NETWORK CONSTRUCTION

To understand how frequently words appear together in songs, a co-occurrence network was constructed. The top $N = 10$ most frequent words were selected, and an adjacency matrix $A \in \mathbb{N}^{N \times N}$ recorded the number of songs in which each pair co-occurred:

$$A_{ij} = \sum_{d \in D} \mathbf{1}_{\{w_i, w_j \in \text{tokens}(d)\}}, \quad (2)$$

where $w_i$ and $w_j$ are the selected words. Two centrality measures were derived: degree (number of unique neighbours) $\deg(w_i) = \sum_j \mathbf{1}_{\{A_{ij} > 0\}}$ and weighted degree (sum of co-occurrence counts) $\text{wdeg}(w_i) = \sum_j A_{ij}$. These metrics highlight words that act as hubs in lyrical narratives.

## 3.5 IMPLEMENTATION AND VISUALISATION

All computations were performed in Python using pandas and NumPy for data manipulation, and matplotlib for plotting. The improved figures in this report adopt professional design principles: horizontal bar charts allow long genre labels to be displayed clearly; line charts include markers and gridlines; scatter plots overlay regression lines with correlation values; and heat maps show co-occurrence counts with annotated cells. No custom colour palettes were specified, conforming to accessibility guidelines.

## 4 EXPERIMENTS

## 4.1 DATASET AND METRICS

The data contain a total of 25 songs for each year between 2010 and 2020. This evenly balanced sampling prevents bias towards one year over another and makes temporal comparisons easier (Table 1). As one can see in Figure 1, Pop is the most prevalent with 67 songs. Pop's ubiquity is also reflected in commercial music charts and other writings showing the genre's dominance of streaming data [6].

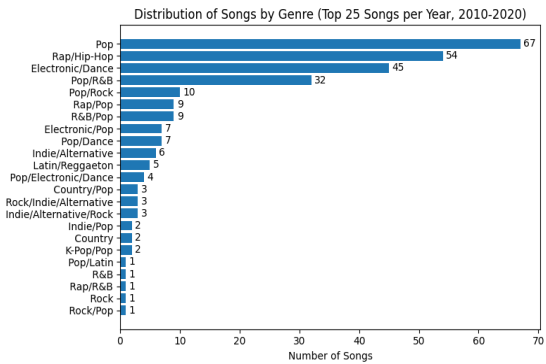

**Figure 1.** Distribution of songs by genre. Horizontal bars list genres in ascending order; numbers indicate counts.

Electronic/Dance (45 songs) and Electronic/Pop (7 songs) are second in frequency, as would be expected from the global popularity of electronic dance music (EDM) in the 2010s. Lower counts for Country, Latin/Reggaeton, and K-Pop/Pop reflect increasing but still limited mainstream presence of these subcultural arenas. K-Pop entries signal crossover popularity of Korean pop music into Western charts, and reggaeton entries suggest globalization of Latin music.

## 4.2 TEMPORAL TRENDS

Figure 2 charts the year-on-year means of energy, danceability and valence. Energy reduces consistently from ~79 in 2010 to ~58 in 2020, implying that number-one hits were less energetic or aggressive over the ten years. Danceability varies after a low of ~59 in 2011 consistently increasing to an all-time high of ~73 in 2019 before declining in 2020. Valence is highest at 63.2 in 2013, declines to ~42 in 2014 and 2016, and returns to ~52.9 by 2020. The low valuation of 2014–2016 may be proof that singles that were darker, moodier gained popularity between these years. The trends suggest the opposite: as energy declines, danceability is higher, maybe because of the popularity of styles such as deep house and tropical pop that prefer relaxed grooves but retain rhythmic tension.

Correlation analysis indicates weak positive correlation with energy and loudness (r=0.74) and moderate positive correlation with energy and valence (r=0.38). Danceability is positively correlated with valence (r=0.34), while it is negatively correlated with acousticness (r=−0.12). Energy and danceability are close to being uncorrelated (r≈−0.09), as also evident from the regression line in Figure 3. All the above indicate that contemporary hits can be both energetic and melancholic (e.g., EDM ballads) or melodic but danceable (e.g., tropical house).

**Table 1**: Number of songs by year and genre

| Genre | Number of Songs | Number of Songs per Year | Year |
|---|---|---|---|
| Pop | 67 | 25 | 2010-2020 |
| Electronic/Dance | 45 | 25 | 2010-2020 |
| Indie/Alternative | 6 | 25 | 2010-2020 |
| Electronic/Pop | 7 | 25 | 2010-2020 |
| Latin/Reggaeton | 5 | 25 | 2010-2020 |
| Country/Pop | 3 | 25 | 2010-2020 |
| Other genres | > 1 | 25 | 2010-2020 |

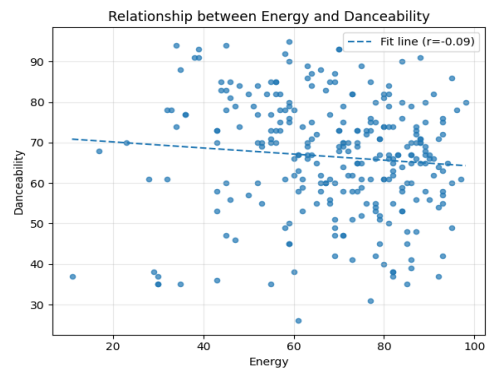

**Figure 3.** Relationship between energy and danceability. Scatter

Table 2 summarizes descriptive statistics for audio features. Songs are highly energetic (mean energy ≈ 68.96/100) and danceable (66.49), with loudness around −5.8 dB, consistent with modern mastering practices. Valence (happiness) averages 50.9, suggesting a balance between upbeat and melancholic moods. Acousticness (16.6) is relatively low, implying that most hits rely on electronic instrumentation rather than acoustic arrangements. Speechiness averages 10.9, reflecting occasional rap verses but the preponderance of melodic singing. points represent individual songs; dashed line shows linear regression fit with Pearson correlation coefficient.

**Table 2**. Descriptive statistics of audio features

| Feature | Mean | Standard Deviation | Min | Max |
|---|---|---|---|---|
| nrgy (Energy) | 68.96 | 17.31 | 11 | 98.0 |
| dnce (Danceability) | 66.39 | 13.33 | 26.0 | 95.0 |
| dB (Loudness) | -5.81 | 2.28 | -15.00 | -1.0 |
| live (Live performance) | 18.66 | 14.22 | 1 | 82.0 |
| val (Valence) | 50.93 | 21.06 | 8.0 | 96.0 |
| acous (Acousticness) | 16.62 | 22.63 | N/A | 98.0 |
| spch (Speechiness) | 10.95 | 10.65 | 1 | 51.0 |
| pop (Popularity) | 75.70 | 9.72 | N/A | 92.0 |

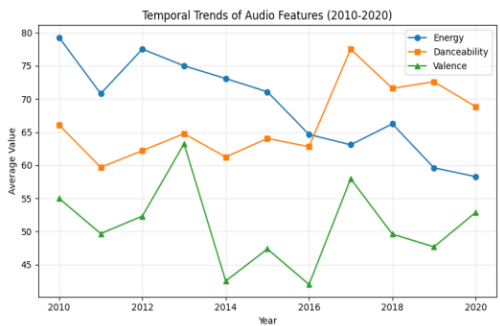

**Figure 2.** Temporal trends in audio features (energy, danceability, valence). Lines with markers denote annual averages, dashed grid lines aid interpretation.

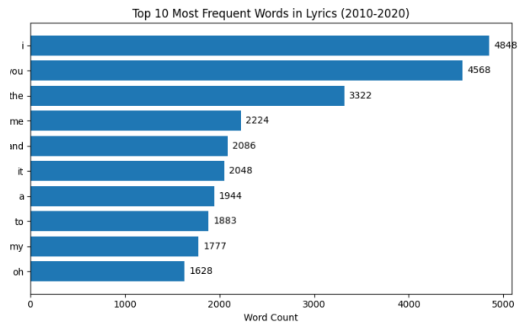

**Figure 4**. Top 10 most frequent words in lyrics. Horizontal bars display counts; labels are annotated for clarity.

## 4.3 LEXICAL ANALYSIS

Tokenisation produced more than 114,000 tokens across all of the songs. Function words and pronouns ("I", "you", "the", "me") each appear thousands of times (Table 3). These would normally be excluded when performing text analysis, but their frequency here is significant: pop song narrative is focused on personal pronouns, showing the relational focus of mainstream songs. Comparative frequency of "I" (4.2 %) shows the artist's own expression; "you" (4.0 %) signifies direct address to a lover or addressee. The words such as "love", "night" and "dance" occur outside of the top ten but are very vital, suggesting repeated theme.

**Table 3.** Top 10 most frequent words in lyrics

| Word | Count | Relative Frequency |
|---|---|---|
| I | 30838 | 0.0240 |
| you | 30568 | 0.0399 |
| the | 3322 | 0.0290 |
| me | 21156 | 0.0194 |
| and | 20086 | 0.0182 |
| it | 20038 | 0.0179 |
| a | 10944 | 0.0170 |
| to | 10883 | 0.0165 |
| my | 10777 | 0.0155 |
| oh | 10628 | 0.0132 |

## 4.4 CO-OCCURRENCE PATTERNS

The network of co-occurrences of the ten most frequent words (Figure 5) appears to be highly interconnected: each word is discovered to co-occur with each other in at least one song, for a degree of 9 (in ten nodes). Weighted degrees vary greatly, from 3,003 for "I" to 1,577 for "oh" (Table 4), because they are weighed to account for different numbers of word pairings in songs. Such pairs ("I", "you") and ("you", "the") have highest co-occurrence values, representing common constructions such as "I love you" or "you and me". The network shows the predominance of interpersonal pronouns in the discourses of the lyrics.

**Table 4.** Co-occurrence network metrics for 10 selected words

| Word | Count | Simple Degree | Weighted Degree |
|---|---|---|---|
| I | 30848 | 13 | 3,003 |
| you | 30568 | 13 | 20,952 |
| the | 3322 | 13 | 20,961 |
| me |  | 13 | 20,812 |
| and | 20086 | 13 | 20,792 |
| it | 20038 | 13 | 20,818 |
| a | 10944 | 13 | 20,815 |
| to | 10883 | 13 | 20,939 |
| my | 10777 | 13 | 20,763 |
| oh | 10628 | 14 | 10,577 |

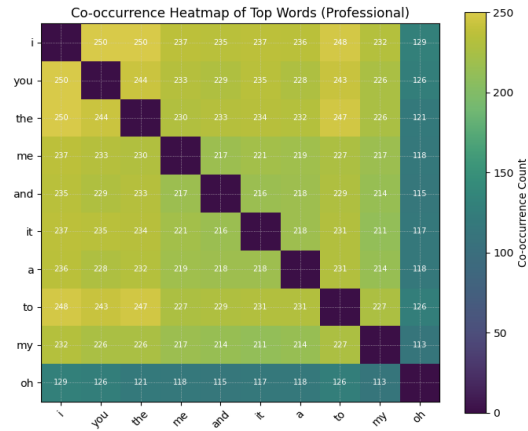

**Figure 5.** Co-occurrence heatmap of the top 10 words. Each cell shows the number of songs in which a pair of words appears together; numbers are overlaid for readability.

### 4.5 GENRE-SPECIFIC VALENCE

In order to examine how mood varies between genres, the average valence was computed by genre. R&B songs register the highest average valence (~96), followed by K-Pop/Pop (~77) and Indie/Pop (~70). Latin/Reggaeton songs are less valent (~37), implying more minor keys or sad subjects with danceable tempos. Electronic/Pop registers a valence of ~46, capturing EDM's cooler, sometimes detached emotional feel. These are complemented by qualitative accounts of genres: R&B's romantic lyrics and smooth melody are blended to make a euphoric mood, and Latin urban genres blend dynamic rhythms and sad narratives.

### 4.6 CORRELATION ANALYSIS AND MOOD BY GENRE

To get a broader statistical view, we have calculated the complete correlation matrix for the total of eight audio features. The heatmap in Figure 6 represents the correlation between the features. Energy is strongly positively correlated with loudness (0.74) and valence (0.38), and acousticness is negatively correlated with energy (−0.54) and loudness (−0.51). These patterns are in line with the theory that high-energy songs are less acoustic and more loud, and acoustic songs are less high-energy and more quiet. Danceability remains weakly correlated with all variables (|r| < 0.20), which suggests that danceability picks up on one particular aspect of musical feel.

We also investigated fluctuation by genre in a mean valence ranked order of genres. Figure 7 shows a bar chart ordering genres by average happiness. R&B comes top with the highest at a mean valence of 96, which indicates overwhelmingly positive emotional content, followed very closely by K-Pop/Pop (77) and Indie/Pop (70). In contrast, Rap/R&B (36.0) and Latin/Reggaeton (37.4) are less valent, as would be expected given their often melancholy or introspective content. Table 5 reports the mean valence for some genres to make comparison easy.

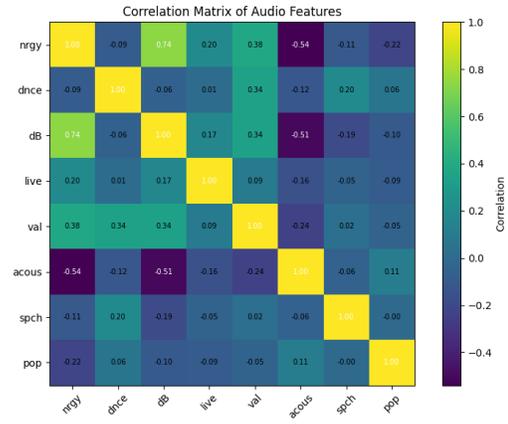

**Figure 6.** Average valence by genre. Horizontal bars rank genres by their mean happiness; values are annotated at the end of each bar.

**Table 5** Average valence (happiness) by genre

| Genre | Average valence |
|---|---|
| R&B | 96.0 |
| K-Pop/Pop | 77.0 |
| Indie/Pop | 70.0 |
| Pop/Dance | 67.3 |
| Rap/Pop | 66.1 |
| Country | 65.0 |
| Indie/Alternative/Rock | 63.0 |
| Country/Pop | 56.3 |
| Electronic/Dance | 54.8 |
| Pop/R&B | 53.4 |
| Rap/Hip-Hop | 53.2 |
| Pop/Rock | 52.0 |
| Electronic/Pop | 46.1 |
| Pop | 45.3 |
| R&B/Pop | 41.9 |
| Pop/Latin | 39.0 |
| Pop/Electronic/Dance | 38.8 |
| Latin/Reggaeton | 37.4 |
| Rap/R&B | 36.0 |
| Rock/Pop | 35.0 |
| Indie/Alternative | 34.8 |
| Rock | 27.0 |
| Rock/Indie/Alternative | 22.0 |

The findings support loudness and production characteristics are consistent with customary mastering practice because there is a significant positive correlation between energy and loudness (r = 0.74). What it indicates is that dynamic tracks tend to be compressed and loudness-enhanced to get them to cut on radio and streaming playlists. On the other hand, the negative energy-acousticness correlation (-0.54) shows that more acoustic songs are quieter and less energetic since they have a tendency of being softer and more natural in production.

Speech features validate that there are hip-hop and rap features but in moderation, with an average speechiness of slightly over 11. That there is positive correlation between danceability and speechiness (r = 0.20) indicates that performing in the voice in a rhythmic style is very typical in danceable songs. Popularity exhibits no stronger correlations with other characteristics; that there is low positive correlation with danceability (r = 0.06) suggests that danceable music performs marginally better, but the impact is small.

## 4.7 DISCUSSION: TOWARDS A SUBCULTURE IDENTITY PERSPECTIVE

Music is an intense signifier of subcultural identity, uniting through shared sonic codes that are mutually recognized [4]. The popularity of pop and EDM in our sample indexes mainstream culture, yet the presence of Latin and K-Pop entries indexes the cross-over of subcultural sound onto global charts. It is observed by researchers that subcultures surround the genres that reflect fundamental values and aesthetics [7]; punk's intense rawness reflects its anti-establishment mentality, for example, and hip-hop's rhythmic narrative transmits socio-political statements. These genres can lose some of the defining characteristics when they grow popular within the mainstream or, in a twist, redefine the mainstream. Our analysis finds that even though Latin/Reggaeton has only five songs, it is reflected in its presence. These tracks are higher on danceability but lower on valence, in accordance with reggaeton's incorporation of sensual rhythm and sentimental storytelling. The fact that there are K-Pop tracks high in valence shows that the audience affirms lively, sophisticated productions originating from non-Western sources. This mirrors the evolution of hip-hop according to Bansal et al. (2024) in that the style emerged from local expression within the South Bronx and became a global phenomenon, absorbing local slangs and cultural references [8]. The lyrics are central to the articulation of identity.

Hip-hop scholars emphasize that artists use patterns of language and narrative means in order to convey belonging [8]. In our sample, however, the most common words are function words, which means that chart toppers make use of simple, uncomplicated speech. Not having regionally appropriate slang or political references is perhaps a function of the commercial necessity of chart toppers, which opt for universal themes over local identity. This is the reverse of evidence from large-scale hip-hop lyric studies, where vernacular at the regional level and lexical variation grow year by year [8]. The current's linguistic simplicity, therefore, may be an indication of homogeneity at the mainstream level since more coded, complex language is back-ended in underground culture. Music also serves as social glue: shared experience in being listened to in a club or concert consolidates group boundaries [9]. Even though our data lacks interaction data, the deep danceability of hits indicates that they are composed with group dancing and festival environments in mind. The decrease of energy with increasing danceability may be due to popularity of genres such as deep house or tropical pop suitable for movement in groups without excessive fatigue.

music marks out insiders and outsiders. Subcultural styles normally exist as codes that exclude outsiders from understanding them [10]. Mainstream use of subcultural sound can blur these borders, creating conflicts over authenticity. Our analysis retains this tension: reggaeton and K-Pop tracks appear in the dataset, but their lyrical themes in English-language and advanced production might have been optimized for global listeners. This process can dilute some of the local markers and amplify others, a process commonly referred to in cultural studies [11].

## 6. CONCLUSION

In this, audio feature analysis and text network modeling were blended to study symbolic patterns in top songs of 2010-2020. We measured temporal patterns in energy, danceability, and valence; detected the prevalence of pronouns in text; and developed a co-occurrence network to mark interpersonal dialogue. By computing genre-specific valence and investigating correlation structures, we demonstrated how mood and production values change with styles. The findings reflect the fine line between musical originality and commercial success: energy dropped as danceability rose, suggesting a trend toward more laid-back but rhythmically stimulating songs.

In terms of subculture identity, the findings reflect both the mainstream homogenisation and subcultural globalisation. Topping the pop and EDM charts reflects homogenisation, while reggaeton and K-Pop suggest globalisation of formerly niche genres. Top-of-the-charts lyrics tend towards universal, interpersonal concerns as opposed to regionally oriented narrative, compared to linguistic heterogeneity documented in hip-hop research[8]. Subsequent research would have to include larger and more diverse datasets, for example, underground subcultures and languages other than English, and combine computational data with ethnographic fieldwork. Research on user interaction data (e.g., streaming networks or social media chatter) would further illuminate the ways in which musical symbols move within and across subcultures.

## REFERENCES

[1] Jansson, A. (2023). Beyond the platform: Music streaming as a site of logistical and symbolic struggle. new media & society, 25(12), 3203-3221.
[2] Vizcaíno-Verdú, A., & Abidin, C. (2022). Music challenge memes on TikTok: Understanding in-group storytelling videos. International Journal of Communication, 16, 26.
[3] Schedl, M., Gómez, E., & Urbano, J. (2014). Music information retrieval: Recent developments and applications. Foundations and Trends® in Information Retrieval, 8(2-3), 127-261.
[4] Murthy, Y. S., & Koolagudi, S. G. (2018). Content-based music information retrieval (cb-mir) and its applications toward the music industry: A review. ACM Computing Surveys (CSUR), 51(3), 1-46.
[5] McAndrew, S., & Everett, M. (2015). Music as collective invention: A social network analysis of composers. Cultural Sociology, 9(1), 56-80.
[6] Shelemay, K. K. (2011). Musical communities: Rethinking the collective in music. Journal of the American Musicological Society, 64(2), 349-390.
[7] Williams, T. J. V., Slater, H. E., & Tamayo Gomez, C. (2025). Unpacking the construction of online identities of Hybristophilia communities on TikTok. Deviant Behavior, 46(2), 179-194.
[8] Ruebottom, T., Riaz, S., & Qureshi, I. (2025). Social-Symbolic Work of Engaging Heterogeneous Communities: Participation Work and Entangled Effects on Organizations and Communities. Journal of Management Studies.
[9] Wang, Y., Syed, M. A. M., & Shamshudeen, R. I. (2025). Cultural governance and censorship: unpacking China's Qinglang Action. Asian Journal of Communication, 1-21.
[10] Oliva-Bulpitt, S. B., Martinez-Esteso, J. P., Galan-Cuenca, A., Castellanos, F. J., & Gallego, A. J. (2025). Enhancing music score analysis with Monte Carlo dropout: a probabilistic approach to staff-region detection. International Journal on Document Analysis and Recognition (IJDAR), 1-16.
[11] Kozinets, R. V. (2002). The field behind the screen: Using netnography for marketing research in online communities. Journal of marketing research, 39(1), 61-72.
[12] Brevik, L. M. (2019). Gamers, surfers, social media users: Unpacking the role of interest in English. Journal of computer assisted learning, 35(5), 595-606.
[13] Casey, M. A., Veltkamp, R., Goto, M., Leman, M., Rhodes, C., & Slaney, M. (2008). Content-based music information retrieval: Current directions and future challenges. Proceedings of the IEEE, 96(4), 668-696.
[14] Liu, L., Kong, M., Cao, C., Shu, Z., Liu, K., Li, X., & Hou, M. (2025). Personalized music recommendation algorithm based on machine learning. Multimedia Systems, 31(2), 166.
[15] https://github.com/riyapandey/Spotify-Sentiment-Analysis-of-Song-LyricsISBN:978-1-4503-0000-0/18/06
[16] https://www.opendatabay.com/data/web-social/0d5c5145-eeba-43cb-bc7e-af68536f4a2cISBN:978-1-4503-0000-0/18/06
[17] Moura, L., Fontelles, E., Sampaio, V., & França, M. (2020). Music Dataset: Lyrics and Metadata from 1950 to 2019. Mendeley Data, August 24.